\documentclass[12pt,preprint]{aastex}
\slugcomment{2014.1 v4}

\shorttitle{Wind braking for the anti-glitch of 1E 2259+586} \shortauthors{H. Tong}

\begin{document}

\title{Anti-glitch of magnetar 1E 2259+586 in the wind braking scenario}

\author{H. Tong\altaffilmark{1}}

\altaffiltext{1}{Xinjiang Astronomical Observatory, Chinese Academy of Sciences, Urumqi, Xinjiang,
    China\\ {\it tonghao@xao.ac.cn}}

\begin{abstract}
The anti-glitch of magnetar 1E 2259+586 is analyzed theoretically.
An enhanced particle wind during the
observational interval will taken away additional rotational energy of the neutron star. This will result
in a net spin-down of the magnetar, i.e., an anti-glitch. In the wind braking scenario of
anti-glitch, there are several predictions: (1) A radiative event will always accompany the anti-glitch,
(2) Decrease/variation of braking index after anti-glitch, (3) Anti-glitch is just a period of enhanced spin-down.
If there are enough timing observations, a period of enhanced spin-down is expected instead of
an anti-glitch.
Applications to previous timing events
of SGR 1900+14, and PSR J1846$-$0258 are also included. It is shown that current timing events of
1E 2259+586, SGR 1900+14, and PSR J1846$-$0258 can be understood safely in the wind braking model.
The enhanced spin-down and absence of an anti-glitch before the giant flare of SGR 1806$-$20 is
consistent with the wind braking scenario.
\end{abstract}


\section{Introduction}

Magnetars are neutron stars powered by their super strong magnetic fields (Duncan \& Thompson 1992).
Timing of magnetars can tell us the strength of their dipole fields,
the activities of their magnetospheres, and the interior of magnetars (Kouveliotou et al. 1998;
Kaspi et al. 2003; Tong et al. 2013).
Glitches are observed from many magnetars. Some of these glitches are associated
with the star's magnetospheric activities (Kaspi et al. 2003). Some of them are not associated with magnetospheric
activities (Dib et al. 2008). These glitches may due to internal angular momentum exchange of the magnetars (Dib et al. 2008).
However, no definite conclusion can be said at present.

Recently, the magnetar 1E 2259+586 showed a sudden spin-down, i.e., anti-glitch (Archibald et al. 2013).
According to Archibald et al. (2013), the anti-glitch of magnetar 1E 2259+586 has the following characteristics.
\begin{enumerate}
 \item Between 2012 April 14 and 2012 April 28, there is a net decrease of the star's spin frequency
 $\Delta \nu \approx -2\times 10^{-7} \,\rm Hz$. This event is dubbed as ``anti-glitch''. After the anti-glitch,
 the star's spin-down rate is about 2-3 times larger (The exact value depends on the timing model employed, model 1 or model 2
 in Archibald et al. 2013). After some 50-90 days, a second timing event may be required, either
 glitch or anti-glitch.  The spin-down rate there has almost returned to its pre-anti-glitch
 level (the net spin-down rate is only 10-20 percent higher).

 \item At the epoch of anti-glitch, the source's $2-10 \,\rm keV$ flux is 2 times higher. Its spectrum becomes harder.
 There is also moderate change in the pulse profile. The flux decreases in a power law form. On 2012 April 21, a short burst
 is observed by Fermi Gamma-ray Burst Monitor.
\end{enumerate}
If this anti-glitch is originated from the magnetar interior,
it may require rethinking of the glitch model of all neutron stars (Archibald et al. 2013). However, there
is also possibility that the anti-glitch is due magnetopsheric activities (as can be seen that it is accompanied
by radiative changes).

During timing studies of pulsars and magnetars, the magnetic dipole braking assumption is often employed (Kouveliotou et al. 1998).
The real case must be some kind of particle wind (a mixture of particles and electromagnetic fields, Xu \& Qiao 2001; Kramer et al. 2006;
Harding et al. 1999; Thompson et al. 2000; Tong et al. 2013).
A particle wind may be generated during the magnetic field decay of magnetars.
Since this particle wind is originated from magnetic field
energy release, its luminosity may also vary significantly as that of the magnetar's X-ray luminosity.
When the wind luminosity is much higher, it will cause a period of enhanced spin-down.
After some time, it may cause an observable spin-down of the magnetar, i.e., anti-glitch.
In our opinion, the anti-glitch of magnetar 1E 2259+586 may be explained in this
wind braking scenario\footnote{During the preparation precess of this paper, we know the work of Lyutikov (2013)
which provides another magnetospheric origin to the anti-glitch of 1E 2259+586. Our calculation is in the
wind braking model of magnetars, while Lyutikov (2013) is from a different point of view.}.

Internal origin of anti-glitch is discussed in Section 2. Anti-glitch in the wind braking scenario is explored
in Section 3. Discussions and conclusions are given in section 4.

\section{Anti-glitch in the internal origin scenario}

At the epoch of the anti-glitch of magnetar 1E 2259+586, its radiative properties (flux, spectrum, and pulse profile) are
also changed. Therefore, internal exchange of angular momentum (which cause the observed anti-glitch) must at the same time trigger
magnetospheric activities.

After the anti-glitch, the spin-down rate is 2-3 time higher. After glitches in normal pulsars, the spin-down rate
is about $1\%$ higher (Link et al. 1992). This is consistent with that the moment of inertia of the crust neutron superfluid
is only about $1\%$ of the star's total moment of inertia (Link et al. 1992). Even considering detailed interaction
of neutron superfluid and the crust, the required moment of inertia of the crust neutron superfluid is only $4-6\%$
(Anderson et al. 2012). Therefore, anti-glitch must involve angular momentum reservoir besides the crust superfluid
(e.g. core superfluid, Page 2012).

Normal spin-up glitches are also observed in 1E 2259+586 in the year 2002 (Kaspi et al. 2003) and 2007 (Icdem et al. 2012).
If these two glitches also originate from internal angular momentum exchange, then it means that the superfluid
component rotates faster than the crust in 2002 and 2007, while in 2012 it rotates slower than the crust.

Timing of magnetars find many glitches, while the anti-glitch event is very rare (previously there is only a candidate anti-glitch
in SGR 1900+14, Woods et al. 1999). In normal pulsars, hundreds of glitches are observed (Esponiza et al. 2011; Yu et al. 2013).
While no anti-glitch is observed in normal pulsars. If anti-glitch in magnetar 1E 2259+586 is
related to its magnetar field (the surface dipole field of 1E 2259+586 is $\le 6\times 10^{13} \, \rm G$, Tong et al. 2013),
then there is also possibility that similar criterion is reached in normal pulsars. Current studies support the idea that
pulsars and magnetars are a unified population (Perna \& Pons 2011). Although the occurrence of an anti-glitch
in normal pulsars may be low, considering that the pulsar population is 100 times larger than the magnetar population,
the possibility of finding one anti-glitch in normal pulsars may be very high.

The high magnetic field pulsar PSR J1846-0258 is a transition object between normal pulsars and magnetars (Gavriil et al. 2008).
There is over recovery after one large glitch in PSR J1846-0258 (coincide with the star's burst activities,
Livingstone et al. 2010). If there is no timing observations
just near the glitch epoch, then this timing event will also be a net spin-down of the pulsar, i.e. anti-glitch. This observation
may tell us that a normal pulsar can also have anti-glitch only when it is burst active.

In summary, the internal origin of anti-glitch in magnetar 1E 2259+586 can not be ruled out at present.
If the anti-glitch of magentar 1E 2259+586
is from interior of the neutron star, then it will urge rethinking of glitch modeling of all neutron stars
(Archibald et al. 2013). From previous observations of glitches in magnetars (some glitches are not associated with
radiative events), if the anti-glitch is of internal origin, some anti-glitch may also not associated with radiative
event. Future observation of one anti-glitch without radiative event (both in normal pulsars and magnetars)
will support the internal origin of anti-glitch.

\section{Anti-glitch in the wind braking scenario}

\subsection{Description of the wind braking model of magnetars}

Many aspects of the wind braking model can be found in Tong et al. (2013) and references therein.
In the wind braking model of magnetars, the magnetic energy is the ultimate energy reservoir.
The X-ray luminosity of magnetars are mainly from the decay of their multipole field
(Thompson \& Duncan 1996; Pons \& Perna 2011).
 A particle outflow may be generated during the decay of the magnetar field (i.e., particle wind,
Thompson \& Duncan 1996; Harding et al. 1999; Thompson et al. 2000; Tong et al. 2013).
The particle wind luminosity may be as high as the star's X-ray luminosity.
There is an Alfv\'{e}n radius of the outflowing particles, where the magnetic
energy density equals the particle kinetic energy density.
The Alfv\'{e}n radius is determined by both the particle wind luminosity and the star's
surface dipole field. It is related to the rotational energy carried away
per unit outflowing mass. This particle wind may dominate the rotational
energy loss rate of magnetars (Tong et al. 2013). Therefore, there are two kinds of magnetic fields in magnetars:
dipole field and multipole field (Pons \& Perna 2011; Dall'Osso et al. 2012; Tong et al. 2013).
The multipole field controls the energy reservoir.
While the dipole field mainly participate in the braking torque.

Some seismic activities may trigger the outburst of 1E 2259+586,
e.g., Alfv\'{e}n wave injection (Thompson \& Duncan 1996),
or field twisting and subsequent untwisting (Beloborodov 2009), etc.
Its X-ray luminosity increases and latterly decays.
The particle wind luminosity also increases during the outburst.
This will caused a period of enhanced spin-down. Because of the sparse of X-ray timing,
observationally we see the star suffers a net spin-down (i.e., anti-glitch).
The duration of this enhanced spin-down period is a parameter of the current model.
Physically, it is determined by the duration of seismic activities and and subsequent energy dissipation
(which may be similar to the untwisting magnetosphere case, Beloborodov 2009).

The wind braking model of magnetars shares some merits with the twisted magnetosphere model
(Thompson et al. 2002), which is employed to build the corona-mass-ejection-like model for anti-glitch (Lyutikov 2013).

\subsection{Anti-glitch of magnetar 1E2259+586 in the wind braking scenario}

From 2012 April 14 to 2012 April 28, the net spin-down of 1E 2259+596 is
$\Delta \nu \approx -2\times 10^{-7} \,\rm Hz$. The corresponding rotational energy loss of the
central neutron star is\footnote{The total rotation energy of magnetar 1E 2259+586 is about
$4\times 10^{44} \,\rm erg$.}
\begin{equation}
 -\Delta E_{\rm rot} = 4\pi^2 I \nu |\Delta \nu|
 =1.1\times 10^{39} I_{45} \left( \frac{|\Delta \nu|}{2\times 10^{-7} \,\rm Hz} \right) \,\rm erg,
\end{equation}
where $I$ is the moment of inertia which has been spun-down during the anti-glitch, $I_{45}$ is the corresponding
moment of inertia in units of $10^{45} \,\rm g \,cm^2$. If the whole neutron star has been spun-down during the
anti-glitch, then $I_{45} \approx 1$. Meanwhile, if only the neutron star crust has been spun-down during the anti-glitch,
then $I_{45} \approx 10^{-2}$. The rotational energy carried away by a particle wind per unit time
is\footnote{Equation (13) in Tong et al. (2013). Here only the simplest case is considered. Detailed consideration of the
wind luminosity only result in quantitative changes (Section 3.3 in Tong et al. 2013).}
\begin{equation}\label{Edotw}
 \dot{E}_{\rm w} = \dot{E}_{\rm d}^{1/2} L_{\rm p}^{1/2},
\end{equation}
where $\dot{E}_{\rm w}$ is the rotation energy loss rate due to a particle wind,
$\dot{E}_{\rm d}$ is the magnetic dipole rotational energy loss rate,
$L_{\rm p}$ is particle wind luminosity (see Section 3.2 in Tong et al. 2013 for details).
During a 14 days interval, the X-ray luminosity after a magnetar outburst can not change too much (Rea \& Esposito 2011).
Therefore, the particle wind luminosity is also expected not to change too much\footnote{Even if the particle wind luminosity changes significantly,
we can average its effect during the corresponding time interval.} (except for small time scale fluctuations,
which may contribute to the large timing noise during magnetar outburst). Therefore, for a constant particle wind, the rotational energy
carried away by the particle wind during a time interval $\Delta t$ is
\begin{equation}
 \Delta E_{\rm rot,w} = \dot{E}_{\rm d}^{1/2} L_{\rm p}^{1/2} \Delta t = \dot{E}_{\rm d}^{1/2} \Delta E_{\rm w}^{1/2} \Delta t^{1/2},
\end{equation}
where $\Delta E_{\rm w}$ is total energy of the outflowing particle wind.
For the anti-glitch of 1E 2259+586, if the particle wind lasts for 14 days
(from 2012 April 14 to 2012 April 28), then $\Delta t \approx 1.2\times 10^6 \,\rm s$. If the particle wind
lasts only for one to two days, then $\Delta t \approx 1.2\times 10^{5} \,\rm s$.
For a constant amount of outflowing particle wind energy $\Delta E_{\rm w}$,
the rotational energy loss is proportional to $\propto \Delta t^{1/2}$. This means that long term particle wind is more effective
in spinning down the neutron star\footnote{This is because, at fixed $\Delta E_{\rm w}$, when $\Delta t$ is larger, the corresponding
Alfv\'{e}n radius will be larger. The rotational energy carried away will also be larger (Thompson et al. 2000; Tong et al. 2013).}.

If the anti-glitch of magnetar 1E 2259+586 is caused by a period of enhanced spin-down, by equating
$-\Delta E_{\rm rot}=\Delta E_{\rm rot,w}$, the total energy of the outflowing particle wind is
\begin{equation}
 \Delta E_{\rm w} =1.4\times 10^{41} I_{45}^2 b_0^{-2} \left( \frac{|\Delta \nu|}{2\times 10^{-7} \,\rm Hz} \right)^2
		    \left( \frac{1.2\times 10^6 \,\rm s}{\Delta t} \right) \,\rm erg,
\end{equation}
where $b_0$ is star's polar dipole magnetic field in units of quantum critical field $4.4\times 10^{13} \,\rm G$.
The correspond particle wind luminosity is
\begin{equation}
 L_{\rm p} =1.1\times 10^{35} I_{45}^2 b_0^{-2} \left( \frac{|\Delta \nu|}{2\times 10^{-7} \,\rm Hz} \right)^2
		    \left( \frac{1.2\times 10^6 \,\rm s}{\Delta t} \right)^2 \,\rm erg \,s^{-1}.
\end{equation}
For 1E 2259+586, if the persistent particle wind luminosity is not very high (compared with the star's rotational energy loss rate
$|\dot{E}_{\rm rot}| =5.5\times 10^{31} \,\rm erg \,s^{-1}$),
then magnetic dipole braking is correct to the lowest order approximation and $b_0 \sim 1$
(the effect of particle wind will be mainly in higher order timing results). On the other hand,
if the persistent particle wind luminosity is much higher than the rotational energy loss rate,
then the magnetic dipole braking is incorrect even to the lowest order approximation and we must
employ the full formalism of wind braking of magnetars (Tong et al. 2013). In this case, $b_0 \sim 0.1$.
Therefore, there are at least two possibilities for the anti-glitch of 1E 2259+586.
\begin{itemize}
 \item \textbf{Case I}, spin-down of the whole neutron star, $I_{45} \approx 1,\ b_0 \sim 1, \ \Delta t \approx 1.2\times 10^{6}\,\rm s$.
 The whole neutron star is spun-down during the anti-glitch. The star's dipole magnetic field is relatively high (can be treated
 as magnetic dipole braking during the persistent state). The particle wind lasts for about 14 days. The total energy
 of the outflowing particle wind is $\Delta E_{\rm w} \sim 10^{41} \,\rm erg$. The corresponding wind luminosity is
 $L_{\rm p} \sim 10^{35} \,\rm erg \, s^{-1}$, which is about five times the persistent soft X-ray luminosity (Zhu et al. 2008).
 Note also that the soft X-ray luminosity just after the anti-glitch is about 2 times the persistent luminosity (Archibald et al. 2013).
 Therefore, during the 14 days before the anti-glitch a higher X-ray luminosity and particle wind luminosity is not impossible.

 \item \textbf{Case II}, spin-down of the neutron star crust, $I_{45} \approx 10^{-2},\ b_0 \sim 0.1, \ \Delta t \approx 1.2\times 10^{5}\,\rm s$.
 The strong particle wind lasts only for one to two days and only the neutron star crust is spun-down during the anti-glitch.
 The star's dipole magnetic field is relatively low (wind braking during the persistent state). The total energy and
 luminosity of the particle wind are $10^{40}\,\rm erg$ and $10^{35} \,\rm erg \, s^{-1}$, respectively.
\end{itemize}
After the anti-glitch, the rotational angular velocity lag between the neutron superfluid and crust will be larger.
This may trigger a subsequent glitch of the neutron star (Anderson \& Itoh 1975). This corresponds to model 1 in
Archibald et al. (2013). After the anti-glitch, the particle wind luminosity may be also variable (decreases and fluctuates
as that of the X-ray luminosity). Therefore another spin-down event (i.e. anti-glitch) of the neutron star may also be caused by the
particle wind after the first anti-glitch. This may corresponds to the model 2 in Archibald et al. (2013).
Meanwhile, the fluctuation of the particle wind will cause a higher level of timing noise. This may also contribute to the uncertainties
in observations. We can not tell whether model 1 or model 2 (Archibald et al. 2013) is more likely from the theoretical point
of view. Both of them are possible in the wind braking scenario.

During the 14 days interval, the average spin-down rate is
\begin{equation}
 \dot{\nu}_{\rm ave,outburst} = -1.7\times 10^{-13} \left( \frac{|\Delta \nu |}{2\times 10^{-7} \,\rm Hz} \right)
 \left( \frac{1.2\times 10^6 \,\rm s}{\Delta t} \right) \,\rm Hz \, s^{-1}.
\end{equation}
For case I, the spin-down rate during the 14 days interval is 17 times the persistent state spin-down rate.
This means that the particle wind luminosity is about 300 times stronger than the persistent state
(from Equation (\ref{Edotw})). Therefore, in case I, the persistent state particle wind luminosity is
about $10^{32} \,\rm erg \,s^{-1}$, the same order as the star's rotational energy loss rate. This
is why it can be treated as magnetic dipole braking during the persistent state (to the lowest order).
For case II, the spin-down rate during the particle wind outburst (which last for one to two days)
is about 170 times the persistent state spin-down rate.
However, only the neutron star crust is spun-down.
Therefore, the particle wind luminosity is just 3 times the persistent state level. The
consequent persistent particle wind luminosity is about $10^{34} \,\rm erg \,s^{-1}$,
the same order as the star's X-ray luminosity (much larger than the rotational energy loss rate).
The star is wind braking during the persistent state. The results here are consistent
with the above calculations based on rotational energy losses.

From Equation (\ref{Edotw}), if the particle wind luminosity after the anti-glitch is five
times the persistent state, then the spin-down rate will change by a factor of
two to three\footnote{The exact value may depends on different modeling of the particle wind.
See Section 3.2 and 3.3 in Tong et al. (2013) for different modeling of particle wind luminosity.}.
This may account for the enhanced spin-down rate
after the anti-glitch. When the X-ray luminosity decreases and returns to its persistent level, the particle wind luminosity
may also decreases and relaxes to its persistent level. After the second event, the spin-down rate has almost relaxed
to its pre-anti-glitch level. This is also the natural result in the wind braking scenario.

When the particle wind luminosity is higher, the corresponding particle number density will be higher.
The magnetospheric scattering optical depth will be larger. This will result in a harder spectrum during
the outburst (Thompson et al. 2002). And the pulse profile may also undergo some changes since the
global magnetospheric structure has now been changed. This is in agreement with spectral
observations of 1E 2259+586 (Archibald et al. 2013).

The soft X-ray luminosity just after the anti-glitch is 2 times the persistent level. Assuming the
soft X-ray luminosity during the 14 days interval is 5 times the persistent state level, the
total outburst energy (during the 14 days interval) is about $10^{41} \,\rm erg$. It is the same order
as the total outflowing particle wind energy.

\subsection{Predictions of the wind braking scenario}

If we attribute the anti-glitch of magnetar 1E 2259+586 to an enhanced particle wind, there are several predictions
in this wind braking scenario.
\begin{enumerate}
 \item Anti-glitch is always accompanied by radiative event. It may include burst and outburst,
 spectrum and pulse profile change, spin-down rate variations etc. In the future, if an anti-glitch
 is observed which is not accompanied by any significant radiative event, then the particle wind model
 for anti-glitch can be ruled out. At present, timing events of 1E 2259+586, SGR 1900+14, and
 PSR J1846$-$0258 (Archibald et al. 2013; Woods et al. 1999; Livingstone et al. 2010)
 are all accompanied by radiative events. This is consistent will the wind braking scenario.

 \item At the epoch of anti-glitch, the particle wind luminosity is higher than the persistent level.
 It will not only cause an enhanced spin-down rate, but also cause a different braking index.
 If the star can be treated as magnetic dipole braking, then there will be a net decrease of braking index
 after the anti-glitch\footnote{Wind braking has a smaller braking index than magnetic dipole braking (Tong et al. 2013).}.
  If the star is wind braking during the persistent state, its braking index will be constant if the
  wind luminosity is constant. However, the particle wind luminosity during the outburst is changing with time.
  Therefore, wind braking after the anti-glitch will cause a variation of the braking index.

 In the real case,
 the timing noise during magnetar outburst will dominate the frequency second derivatives.
 Only long after the outburst, if there are still some particle wind relics, then it may cause
 a smaller/variation braking index.
 This may be case of braking index variation of PSR J1846$-$0258 (a smaller braking index, Livingstone et al. 2011).

 \item In the wind braking scenario, the anti-glitch is just a period of enhanced spin-down. If the observations are
 very sparse, then an anti-glitch is expected. However, if enough observations are available, it will be a period of
 enhanced spin-down. This may explain the differences between timing event of 1E 2259+586 and PSR J1846$-$0258
 (Archibald et al. 2013; Livingstone et al. 2010). In the future, if another magnetar shows a period of enhanced spin
 down, then by dropping the early observations, an anti-glitch is expected. This procedure can also be applied to
 previous timing observations of PSR J1846$-$0258 (e.g. Livingstone et al. 2010).

\end{enumerate}

Our wind braking model for anti-glitch and the corona-mass-ejection-like model of Lyutikov (2013)
are both in the magnetospheric domain. If future observations can constrain the time scale for anti-glitch,
e.g., a few seconds (even minutes or some other relatively short time scale), then the wind braking model is not favored, and the corona-mass-ejection-like model may be
a better approximation to the real case.
There is no anti-glitch event in the 2004 giant flare of SGR 1806$-$20 (Woods et al. 2007). This may provide some
challenge to the corona-mass-ejection-like model. Future giant flare and intermediate flare observations of
magnetars without an anti-glitch event will further challenge this model.

Considering that the timing noise will also increase during the outburst, the timing result may be biased to some degree.
However, a radiative event will always accompany the anti-glitch in the wind braking scenario.
Future detection of one anti-glitch without radiative event (both in normal pulsars and magnetars) can rule the
wind braking model (and all magnetospheric models).

\subsection{Applications to previous events of SGR 1900+14 and PSR J1846$-$0258}

The same procedure for the anti-glitch of magnetar 1E 2259+586 can also be applied to the timing events of
magnetar SGR 1900+14 (Woods et al. 1999) and PSR J1846$-$0258 (Livingstone et al. 2010). In the following,
only case I is discussed, i.e. spin-down of the whole neutron star.

\subsubsection{The timing event of SGR 1900+14}

For SGR 1900+14, the star has a net spin-down of $\Delta \nu = -3.7\times 10^{-5} \,\rm Hz$ during 80 days
observational interval (Woods et al. 1999). In order to account for this spin-down event, the total energy and luminosity
of the particle wind are respectively
\begin{equation}
 \Delta E_{\rm w} =4.3\times 10^{44} I_{45}^2 b_0^{-2} \left( \frac{|\Delta \nu|}{3.7\times 10^{-5} \,\rm Hz} \right)^2
		    \left( \frac{7\times 10^6 \,\rm s}{\Delta t} \right) \,\rm erg,
\end{equation}
and
\begin{equation}
 L_{\rm p} =6.2\times 10^{37} I_{45}^2 b_0^{-2} \left( \frac{|\Delta \nu|}{3.7\times 10^{-5} \,\rm Hz} \right)^2
		    \left( \frac{7\times 10^6 \,\rm s}{\Delta t} \right)^2 \,\rm erg \,s^{-1}.
\end{equation}
The surface dipole field of SGR 1900+14 is relatively high, its corresponds to $b_0 \sim 10$ (both in the magnetic dipole braking
and wind braking case, Tong et al. 2013). The corresponding particle wind energy and luminosity are $\Delta E_{\rm w} =4.3 \times 10^{42} \, \rm erg$
and $L_{\rm p} =6.2\times 10^{35} \,\rm erg \,s^{-1}$, respectively. It is higher than that of 1E 2259+586
(the radiative event in SGR 1900+14 is also much stronger). The particle wind luminosity is about 7 times the
star's persistent soft X-ray luminosity (Mereghetti et al. 2006). It naturally results in a spin-down rate
$\sim 2.3$ times the persistent level (Woods et al. 1999).

SGR 1900+14 showed a possible net spin-down before its 1998 giant flare (Woods et al. 1999). However,
 a similar net spin-down is not seen in SGR 1806$-$20 before its 2004 giant flare (Woods et al. 2007).
 Only a period of enhanced spin-down is observed (Woods et al. 2007, Figure 1 there). This is consistent with the wind braking
 model for anti-glitch: there is no anti-glitch. ``Anti-glitch'' is just a period of enhanced spin-down.
 Only when the observations are very sparse, an anti-glitch is expected. If there are enough timing observations,
 a period of enhanced spin-down is expected. The different timing behaviors of SGR 1900+14 and SGR 1806$-$20
 can be understood naturally in the wind braking scenario.

\subsubsection{The timing event of PSR J1846$-$0258}

The timing event of PSR J1846$-$0258 resulted in a net spin-down of the neutron star $\Delta \nu \approx -10^{-4}\,\rm Hz$
(Livingstone et al. 2010). The interval for this timing event is about 240 days (time between the glitch epoch and when
phase coherence is regained). Assuming the spin-down is caused by a particle wind, the energy and luminosity of the
particle wind are respectively
\begin{equation}
 \Delta E_{\rm w} =4.4\times 10^{42} I_{45}^2 b_0^{-2} \left( \frac{|\Delta \nu|}{10^{-4} \,\rm Hz} \right)^2
		    \left( \frac{2\times 10^7 \,\rm s}{\Delta t} \right) \,\rm erg,
\end{equation}
and
\begin{equation}\label{Lp1846}
 L_{\rm p} =2.2\times 10^{35} I_{45}^2 b_0^{-2} \left( \frac{|\Delta \nu|}{10^{-4} \,\rm Hz} \right)^2
		    \left( \frac{2\times 10^7 \,\rm s}{\Delta t} \right)^2 \,\rm erg \,s^{-1}.
\end{equation}
For PSR J1846$-$0258, its dipole field is $b_0 \sim 1$. Therefore, the typical particle wind energy and luminosity
are $\Delta E_{\rm w} =4.4\times 10^{42} \,\rm erg$ and $L_{\rm p} =2.2\times 10^{35} \,\rm erg \,s^{-1}$, respectively.
According to Livingstone et al. (2010), the pre-glitch spinning down trend ($\nu$, $\dot{\nu}$, and $\ddot{\nu}$) is removed
before all the analysis there. Therefore, the above spin-down $\Delta \nu \approx -10^{-4}\,\rm Hz$ is relative spin-down
to the persistent state. The particle wind luminosity in Equation (\ref{Lp1846}) is the variation of the persistent state
particle wind luminosity. The average spin-down rate during this spin-down event is just the variation of the persistent state
spin-down rate, about $7\% \left(\frac{|\Delta \nu|}{10^{-4} \,\rm Hz}\right)  \left(\frac{2\times 10^7 \,\rm s}{\Delta t} \right)$.
The rotational energy loss rate of PSR J1846$-$0258 is about $8\times 10^{36} I_{45}\,\rm erg \,s^{-1}$.
Therefore, the variation of particle wind luminosity is just several percent of the rotational energy loss rate:
\begin{equation}\label{Lpfraction}
L_{\rm p}/\dot{E}_{\rm rot} = 2.7\% \, I_{45} b_0^{-2} \left( \frac{|\Delta \nu|}{10^{-4} \,\rm Hz} \right)^2
				\left( \frac{2\times 10^7 \,\rm s}{\Delta t} \right)^2
\end{equation}
During the persistent state, PSR J1846$-$0258 is a high magnetic field, rotation-powered pulsar (Livingstone et al. 2006). When there
are some magnetic activities, an additional particle component is generated. This will cause a higher spin-down rate
and result in a net spin-down of the pulsar after some time.

\section{Discussions and conclusions}

Here we attribute the timing events of magnetar 1E 2259+586, SGR 1900+14, and PSR J1846$-$0258
to the enhancement of particle wind. While many aspects of the timing of magnetars may be related with the wind braking.
For example:(1) They have a higher level of timing noise than normal pulsars (Archibald et al. 2008).
(2)Their period derivatives can vary significantly (Dib et al. 2009). (3) The timing of the low magnetic field
magnetars (Tong \& Xu 2012, 2013 and references therein). All these aspects may be related with the wind braking
mechanism of magnetars (Tong et al. 2013).

Anti-glitch (also glitches) is rare in magnetars. There may be many reasons.
(1) Magnetars have a higher timing noise than normal pulsars.
Glitches are mainly found in AXPs, hardly in SGRs. This is due to that SGRs has a higher level of timing noise than AXPs
(Woods et al. 2002).
This may account the rarity of anti-glitches in magnetars, at least partially.
(2) In our wind braking model of anti-glitch, anti-glitch is just a period of enhanced spin-down.
If enough timing is available, a period of higher spin-down rate is expected instead of an anti-glitch.
For example, for SGR 1806$-$20 before its 2004 giant flare, a period of enhanced spin-down is acturally observed
(Figure 1 in Woods et al. 2007). And many spin-down rate variations are seen in magnetars (Dib et al. 2009).

For normal pulsars, their timing and pulse profile variations are correlated (Lyne et al. 2010).
Intermittent pulsars have a higher spin-down rate when the pulsar is on (Kramer et al. 2006).
Therefore, the spin-down rate variations in normal pulsars, PSR J1846$-$0258,
intermittent pulsars, SGR 1900+14, and 1E 2259+586 may be a continuous distribution.
This aspect also indicates that normal pulsars and magnetars are an unified population.

If the anti-glitch of magnetar 1E 2259+586 originates from the interior of the neutron star, then
rethinking of the glitch modeling of all neutron stars is needed (Archibald et al. 2013). Even if the anti-glitch of 1E 2259+586
is of magnetospheric origin, rethinking of glitch modeling of all neutron stars is also needed.
The reason is that some glitch of magnetars are associated with radiative events (Kaspi et al. 2003)
while some are not (Dib et al. 2008). The glitches of normal pulsars are rarely associated with magnetospheric
changes\footnote{In recent years, some magnetospheric changes associated with glitches are discovered.
For example, the braking index of the Crab pulsar changed after its glitches (Wang et al. 2012).}.
Therefore, observationally there are two types of glitches in pulsars and magnetars (Tong \& Xu 2011):
Type I which is radiation-quiet (not associated with radiative events) and Type II which is radiation-loud
(associated with radiative events). This must be explained by future glitch modeling of both pulsars and
magnetars.

In conclusion, the anti-glitch of magnetar 1E 2259+586 (Archibald et al. 2013) can be understood safely
in the wind braking scenario. Future observation of one anti-glitch not accompanied by radiative event
or one anti-glitch with very short time scale can rule out this model.
At present, the timing events of 1E 2259+586, SGR 1900+14, and PSR J1846$-$0258 are
consistent with the wind braking scenario.

\section*{Acknowledgments}
The author would like to thank J. P. Yuan, and R. X. Xu for discussions.
H. Tong is supported by NSFC (11103021),West Light Foundation of CAS (LHXZ201201), Xinjiang Bairen project,
and Qing Cu Hui of CAS.

\end{document}